# SMOOTH OPTICAL SELF SIMILAR EMISSION OF GAMMA-RAY BURSTS.

Vladimir Lipunov[1,2], Sergey Simakov[1], Evgeny Gorbovskoy[1], Daniil Vlasenko[1,2]

1 Lomonosov Moscow State University, Sternberg Astronomical Institute 19234, Russian Federation, Moscow, Universitetsky prospect, 13
2 Lomonosov Moscow State University, Physics Department
lipunov2007@gmail.com

## ABSTRACT

We offer the new type of calibration for gamma-ray bursts (GRB), in which some class of GRB can be marked and has common behavior. We name this behavior Smooth Optical Self Similar Emission (SOS Similar Emission) and identify this subclass of gamma-ray bursts with optical light curves described by a universal scaling function.

## 1. INTRODUCTION

The gamma-ray bursts are among the most distant and powerful phenomena in the Universe. They appear to accompany the formation of black holes when the collapsing matter carries excess angular momentum (Woosley & Heger 2006; Lipunov & Gorbovskoy 2007). This is generally the case during core collapse of massive fast rotating stars (Paczynski 1986) or the merging of neutron stars (Blinnikov et al. 1984). Also GRB can be produced by rapidly rotating magnetars (Usov 1992, Zhang & Meszaros 2001).

We are yet far from fully understanding of the gamma-ray burst (GRB) process, whose study is complicated by a vast variety in the observed behavior of gamma-, x-ray, and optical emission. Therefore the identification of any common patterns would be a great step forward toward the development of a complete theory of the phenomenon.

The most universal pattern observed in gamma-ray bursts is the power law fading of the x-ray and optical afterglow, which continues for several hours/days after the explosion and decreases according to a power-law. This afterglow can be satisfactorily explained by the self-similar propagation of a powerful relativistic shock (Blandford & McKee 1976), which is produced by a point explosion and propagates through the surrounding interstellar medium (Meszaros & Rees 1997; Wijers et al. 1997; Vietri 1997; Katz & Piran 1997; Sari et al. 1998) . However, so far no attempts have succeeded in identifying any general patterns in the initial part of the light curve.

Successful detections of the optical emission simultaneously with the gamma-emission (i.e. prompt optical emission) have proven to be rare since the discovery of the first such coincidence in January 1999 by ROTSE (Robotic Optical Transient Search Experiment) (Akerlof et al. 1999).

Although the discovery is 17 years old fewer than twenty successful optical observations coincident or close-to-coincident with the gamma-ray burst have been made so far, and fewer than ten of them are suitable for our analysis. This is in sharp contrast with the power-law tail of the optical and x-ray afterglow, which has been observed several hundred times. The reason is that GRBs usually last no longer than a few tens or hundreds of seconds. Consequently, prompt and early optical emission of a GRB is much harder to observe than the afterglow.

Previously there was much effort to explore the early optical afterglows (Panaitescu & Vestrand 2008, 2011; Oates et al. 2009; Liang et al. 2013; Wang et al. 2013; Kann et al. 2010). Some of these papers have discussed general trends for the early optical behavior and looked at the implications for the dynamics of the fireball and emission mechanism. It has already been pointed out (Vestrand et al. 2005, 2006) that two types of the behavior of optical flux of gamma-ray bursts are observed. In the first case, the optical emission appears simultaneously with the gamma-ray emission and correlates with it for example GRB080319B (Racusin et al. 2008), GRB100901A (Gorbovskoy et al. 2012). In the second case the optical emission appears before the very end or even after the end of the gamma-ray burst and varies very smoothly: it first increases, reaches a maximum, and then gradually fades into the afterglow. Hereafter we refer to this second, frequently observed type of behavior, as Smooth Optical Self Similar Emission (SOS similar emission). Below we discuss this type of behavior, which shows no particular correlation with the harder gamma-ray emission.

## 2. MASTER OBSERVATIONS

Half of the described GRB were observed at MASTER telescopes in one photometric system and the main goal is to introduce this unique observational results and to demonstrate common behavior for found such types of optical counterparts of GRBs.

MASTER Global Robotic Net (*http://observ.pereplet.ru/* ) consists of new generation of fast robotic telescopes MASTER-II installed on different continents during recent years. MASTER-II is twin 40-cm optical robotic telescopes located in the following observatories: MASTER-Amur(Far east of Russia, Blagoveshchensk), MASTER-Tunka (Russia, Baykal lake), MASTER-Ural (Russian, Ural mountains), MASTER-Kislovodsk (Russia, Caucasus mountains), MASTER-SAAO (South Africa), MASTER-IAC (Spain, Canarias), MASTER-OAFA(Argentina).

MASTER observatories equipped with identical own photometers and controlled by identical software (Lipunov et al. 2010, Kornilov et al. 2012; Gorbovskoy et al. 2013, Lipunov 2016b,c). The identical equipment gives us the possibility to have more than 12 continuous hours observations of optical counterparts of gamma-ray bursts in identical photometric systems. Therefore combining photometric data for different gamma-ray bursts observed from different parts of MASTER Net is a well-justified astronomical process. Taking into account the large field of view of MASTER telescopes (2 x 4 square degrees) we use a large number (3000 to 10000) of reference stars for reduction. As a result, the photometric errors of large catalogs in particular those of USNO-B1 should be leveled out.

The main MASTER unique feature is own software to reduce our wide-field images in real-time and to discover new optical transients in our images online, i.e. within 1-2 minutes after readout from the CCD. This information includes the full classification of optical sources from the image, the data from previous MASTER-Net archive images for every sources, full information from VIZIER database and all open source (for ex., Minor planet checker center), derivation of orbital elements for moving objects, etc. With this software we discover more then 1300 (up to February 2017) optical transients in fully automatic mode. These optical transients are the following: optical counterparts of gamma-ray bursts (including bursts registered by FERMI with extensive error box (Lipunov et al. 2016b), QSO flares, Super Novae, Novae, dwarf novae (3 types), anti-novae (Lipunov et al. 2016a) and another cataclysmic variables, asteroids (including potentially hazardous NEOs (near earth objects)) and comets (Lipunov et al. 2007;

Gorbovskoy et al. 2013). Fast automatic identification is very necessary for fast transients investigation.

Another MASTER key attributes are the fast alert pointing, simultaneous polarization and photometry observations by twin telescopes in parallel and non-parallel mode (MASTER has addition free axis that allows to change the parallel configuration increasing the field of view twice, that is very usefull to observe FERMI and LIGO error-boxes). MASTER is connected to the greate physics experiments: GRB Network (to observe gamma-ray SWIFT, Fermi, MAXI, IPN, etc alerts), LIGO-VIRGO gravitational wave detectors collaboration (Abbott et al. 2016) and to the ANTARES (Dornic et al.2015a,b) and Ice-Cube neutrino observatories to investigate their localization areas for possible optical counterparts and to discover new transients in real-time.

Gamma-ray burst observations are usually made simultaneously with the twin MASTER telescopes equipped with perpendicularly oriented polarizers (Gorbovskoy et al. 2016, Pruzhinskaya et al. 2014). In these cases unfiltered magnitudes are computed from the R- and B-band magnitudes of thousands reference stars (USNO-B1), having been contained at 4 square degrees at each MASTER image, by the formula $m = 0.2B + 0.8R$.

### 3. DATA ANALYSIS

In spite of more than several dozens years of GRB investigation, there are not many GRB counterparts light curves, that both has rising part of LC and are the earlier, because it is very complicated task to get prompt or earlier their observations. We used only the earlier data, including prompt points, that has rising part on the light curve for analysys. This is one of the main advantage of this work. The beginning of the observations (LC) must be close to T90 or earlier.

Our analysis is based on eight good-quality **early** observations of gamma-ray bursts that exhibit the latter behavior.

Five of these observations were observed by the MASTER Global Robotic Net (Lipunov et al. 2010,2016b) and they were the base to obtain the universal light curve. They are the following: GRB100906A (Gorbovskoy et al. 2012), GRB121011A (Pruzhinskaya et al. 2014), GRB140629A (Gorbovskoy et al. 2014) and GRB150413A (Tyurina et al. 2015; Gorbovskoy et al. 2016) . We added GRB080810A (Page et al. 2009) , GRB080710 (Kruhler et al. 2009), GRB071010 (Covino et al. 2008) and GRB060605A (Rykoff et al. 2009) detected by SWIFT to our 4 GRBs, see Figure 1.

All photometry data for the GRBs observed by MASTER obtained at polarizers in white color (unfiltered, the details see Pruzhinskaya et al. 2014).

We analysed all bursts with a maximum at optical light curve described at following works: Panaitescu & Vestrand 2008, 2011; Oates et al. 2009; Liang et al. 2013; Wang et al. 2013; Kann et al. 2010 and took only the bursts that satisfy to the following critheria.

1. Optical light curve has a monotonic rising part (with small fluctuations) and monotonic decay.

The gamma-ray bursts GRB021211, GRB050319, GRB050401, GRB050416, GRB050525, GRB050908, GRB050922C, GRB051109a, GRB051111, GRB060927, GRB061121, GRB061126, GRB071025, GRB071025, GRB060124, GRB060210, GRB060714, GRB060729, GRB060206,

GRB060526, GRB060708, GRB060908, GRB060912 , GRB061021, GRB061007, GRB070529, GRB080310, GRB070318, GRB080928, GRB080129, GRB080928, GRB090423, GRB090510, GRB050730, GRB 110205A do not satisfy of this condition. The GRB070802 (see Zaninoni et al. 2013) was excluded because of late beginning of the optical observations.

2.  Optical light curve must have only one maximum (only one peak). The GRBs that don't satisfy to this condition are the following: GRB050904, GRB060904B, GRB050801, GRB050802, GRB060512, GRB070318, GRB060206, GRB090418, GRB 071031.

3.  Sampling points of the previous maximum must be more than 2. The GRBs that don't satisfy to this condition are the following: GRB070411, GRB050820, GRB050801, GRB060124, GRB061121.

4.  The time of the beginning of GRB ($t_0$) must be close to the trigger time ($t_{trigger}$). It means, that the trigger can be started, for ex., from the precursor, i.e. it can be not the just beginning in every case of registration by gamma-detectors.
( $t_{trigger} - t_{GRB}$ ) << ($t_{max} - t_{trigger}$ )
In the case of uncertainty of the start time or trigger time, such GRB is not considered).

5.  There is no correlation with gamma emission (GRB081008A was excluded by this criteria).

6.  The optical measurements must be unfiltered (white color) or in R band. The GRBs that don't satisfy to this condition are the following: GRB060607 and GRB060418.

The first classical prompt optical emission of GRB990123A detected by ROTSE-I (Akerlof et al. 1999) was excluded because it was detected by another gamma detector - BATSE, which used a triggering system that differs from the SWIFT's one. So only eight bursts (listed above) satisfy to all of these criteria.

The common feature of the light curves in Fig. 1 is their non-monotonic behavior combined with a very smooth light variation. I.e. all light curves have a smooth rise, smooth peak and smooth decay without breaks or jumps passing into each other.

As is evident from these curves, their temporal and photometric properties differ quantitatively from each other with the times and amplitudes of maxima spanning one and two orders of magnitude, respectively.

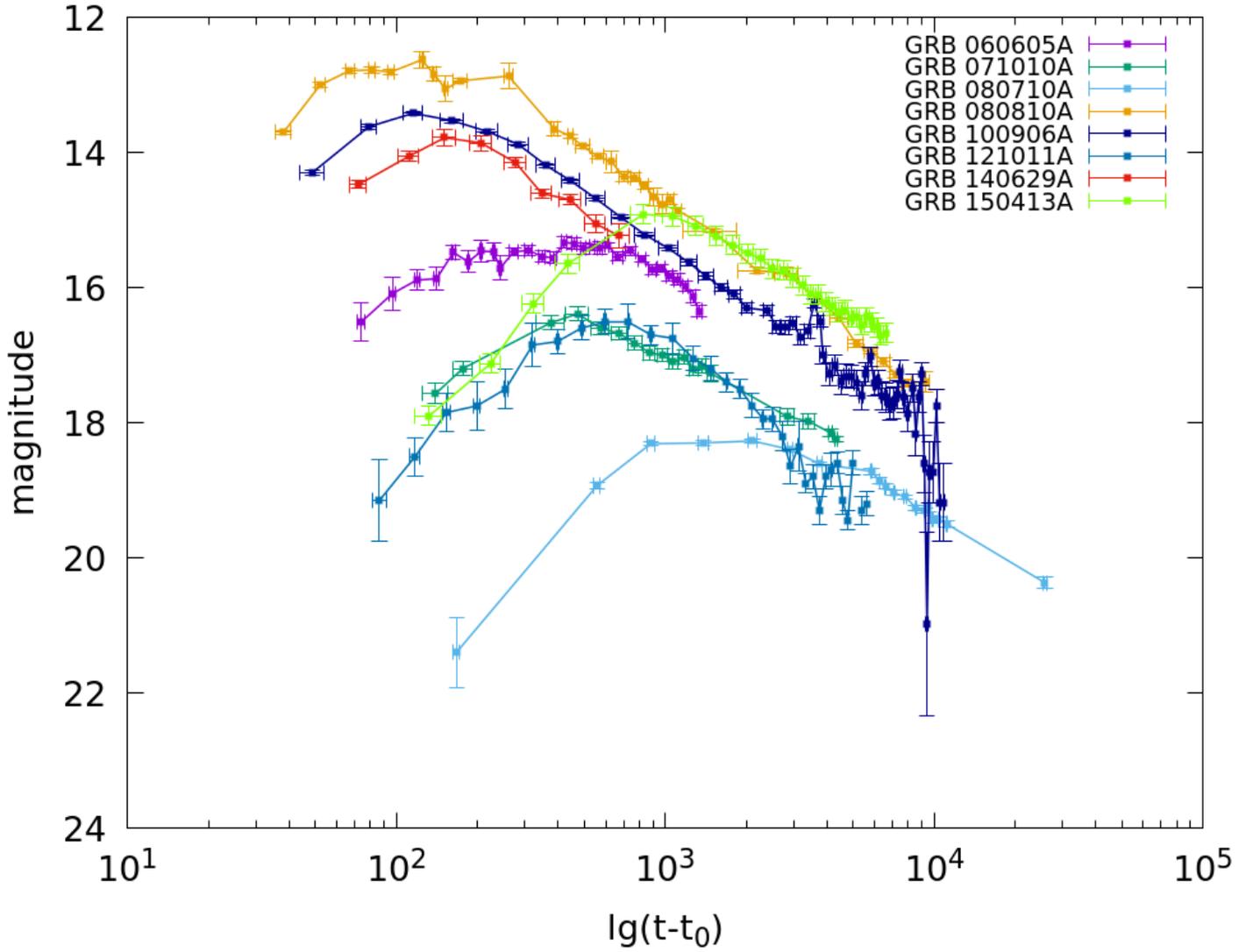

FIGURE 1. There are eight gamma-ray bursts with light curves, that satisfy to our 1-6 CriteriaThe time of the gamma-ray burst beginning it $t_0$. The trigger might work on a precursor or vice versa after beginning of direct burst. However, for all the selected events we use $t_0 = t_{trigger}$ There are the light curves (visual magnitude vs time from detection) of five gamma-ray bursts, observed by the MASTER Global Network, and 4 LCs , observed by other authors: GRB170202A, GRB060605A, GRB080810A, GRB 100906A, GRB121011A, GRB140629A, GRB080710 ,GRB071010 and GRB150413A.

## 4. SOS SIMILAR OPTICAL EMISSION

If we calculate the normalized light curves for GRB050820A, GRB060605A, GRB080810A, GRB100906A, GRB121011A, GRB140629A, GRB150413A, we see the universal shape of the SOS similar emission (Fig. 2). The smooth line is calculated by the approximate formula $m - m_{max} = -2.5 \lg(F/F_{max})$, $\tau = t/t_{max}$. To eliminate the effect of time dilation due to the cosmological redshift, we use dimensionless variable equal to the time elapsed since the onset of the gamma-ray burst divided by the delay of the maximum optical flux, i.e. $\tau = (t - t_0)/(t_{max} - t_0)$. We use R band (Vega system) magnitude for this plot. Following Oates et al. 2009, the peak time $t_{max}$ and $F_{max}$ were determined from a Gaussian fit of each light curve in log time.

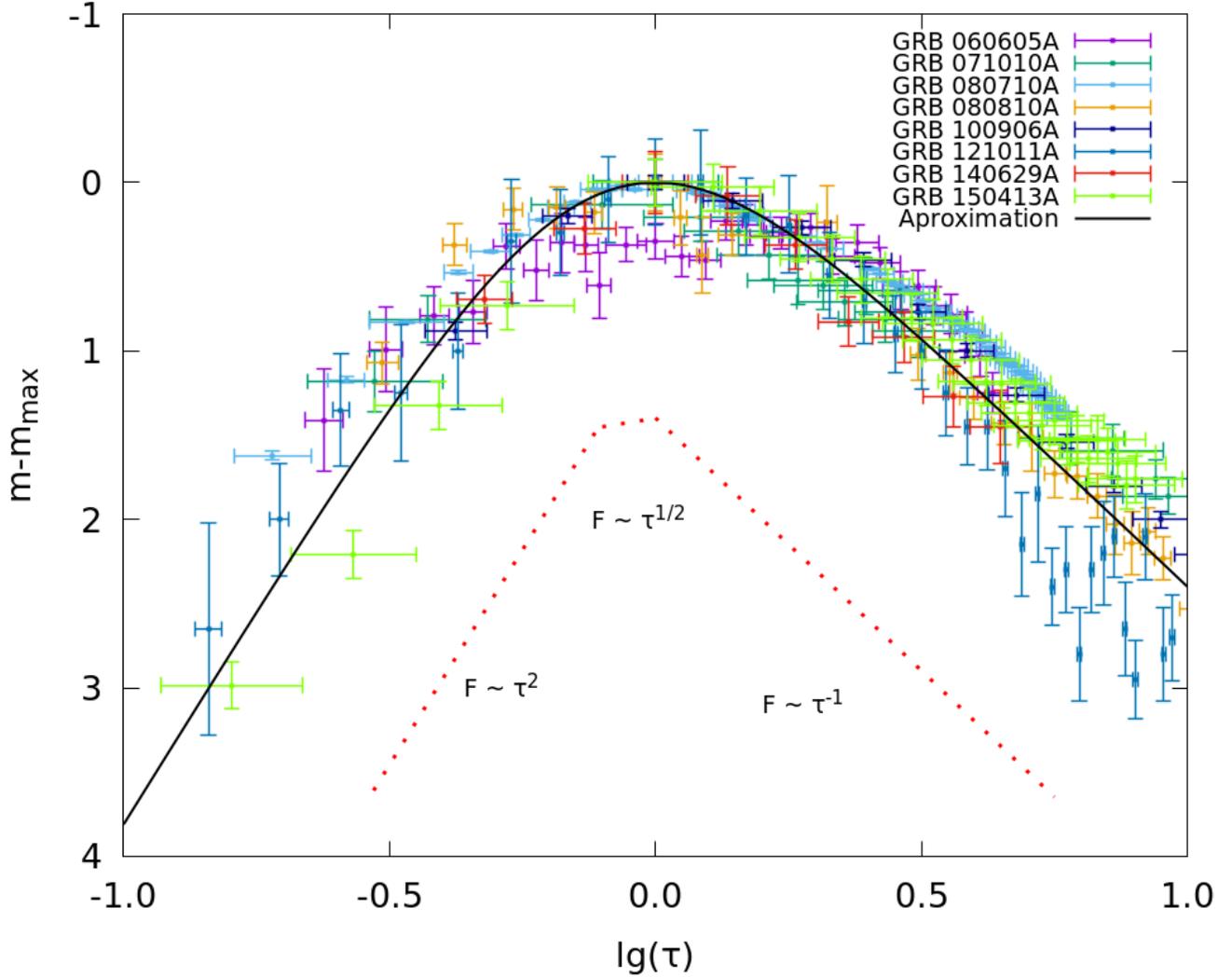

FIGURE 2. The normalized light curves for GRB170202A, GRB050820A, GRB060605A, GRB080810A, GRB100906A, GRB121011A, GRB140629A, GRB150413A. There are eight GRB light curves from Fig.1 in the new coordinates: the brightness is nor-malized to the maximum and the time is normalized to the time of the maximum. The smooth line is given by the approximate formula in the text $m - m_{max} = -2.5 \lg(F/F_{max})$; $\tau = t/t_{max}$.

We nevertheless may combine different gamma-ray bursts into a single plot by introducing modified coordinates, as in Fig.2.

We adopt $t_0 = t_{trigger}$, the trigger action time, almost everywhere. However, care must be taken, e.g., in the cases where observations are triggered by the precursor or when analyzing gamma-ray bursts discovered on different gamma-ray observatories, because of the differences in their trigger systems. That is why we present mostly the gamma-ray bursts discovered by the same Swift BAT observatory. The numerator and denominator depend identically on the redshift and therefore the resulting ratio can therefore be considered to refer to the commoving frame of the gamma-ray burst.

Note that the fact that the smooth variation of the intensity of optical radiation can be described by a universal function is indicative of a universal mechanism and location of the source of this radiation. The fact that this universal function is expressed in terms of some dimensionless arguments suggests that from the very beginning of SOS similar emission we are dealing with a self-similar hydrodynamic propagation of an ultra-relativistic blast wave. We propose to the phenomenological formula:

$$F = F_{max}(\beta \tau^{\beta-1})^\alpha / (1+(\beta-1)\tau^\beta)^\alpha , \qquad (1)$$

where $F$ is the optical flux, $\alpha \sim 1.2$ and $\beta \sim 2.71 \sim e$, $\tau = (t-t_0)/(t_{max} - t_0)$. We put $\beta=e$ in Fig.2. We note that simillar formula proposed as smothed broken power low by Zhong et al. 2016.

In the figure 2 we also depicted a broken power law fitting that illustrates the usual representation of the light curve. It shifted down for convenience. The first link of the broken curve corresponds to the onset stage. The slope of the short link near the maximum is worse determined, but does not contradict to the prediction of self-similar solution $F_\nu \propto \sqrt{t}$ (Gruzinov & Waxman 1999). Decayng part of the curve is fully consistent with the standard model of external forward shock, where light curve changes as power low $F_\nu \propto t^{-1.2}$.

## 5.DISCUSSION

The universal form of the optical light curves found by our analysis agrees well with the predictions of the model of jet or fireball penetrating into the environment (Meszharos & Rees 1997) . On the border of the fireball there is a shock wave compressing the surround matter. Electrons of compressed gas are accelerated to relativistic velocity and give rise to synchrotron emission which is observed as the afterglow. Besides the external forward shock there is a reverse one propagating into the expanding primary shell. It is relatively short – lived but the synchrotron emission of the electrons heated in this wave essentially contributes to the overall flux at early stage of the afterglow. Sometimes this contribution dominates.

The existence of a reverse shock explains (not completely) a wide diversity of morphological features in the light curves. Indeed, the dynamics and emission of the reverse shock depend on whether the ejecta is thick or thin (Kobayashi 2000, Gao et al. 2013), and whether the medium is ISM or wind. (Recall that the shell is thin, if upon reaching its inner border, the reverse shock remains non-relativistic.). In a random sample one can hardly expect the same properties of the fireball. Consequently the light curves of different object with reverse shock should be markedly different (and often have two peaks). As a result, there would be a wide variations on the left side of the chart 2.

In accordance with the above selection rules, we consider only light curves with a single maximum. The unimodality and the lack of broad variations the left of the maximum on the chart 2 suggest that or external shock dominates in luminosity, or the reverse wave "heat but not shine".

The power of the synchrotron emission depends on the spatial density of the electrons and their energy which in turn are determined by blastwave dynamics. When the fireball slow down gaining a mass, i.e. at the stage of the afterglow, the shock wave dynamics is well described by self – similar solution (Blandford & McKee 1976). By a remarkable way hydrodynamical parameters in this case depend just on initial energy of the fireball $E_0$, $\Gamma_0$ - the initial Lorentz factor of the fireball and the density of surrounding matter $n_0$.

The self-similarity is a manifestation of the special internal symmetry of the physical system. It allowes to reduce the number of independent variables, herewith a solution of the hydrodynamic equations depends on a limited number of dimensionless quantities. In the case of adiabatic reverse shock exploding in a homogeneous interstellar medium there are two dimensionless combinations $A \equiv r/ct$ and $B \equiv E_0 t^2/n m_p r^5$.

Since any hydrodynamic quantity can be represented as a function of A and B (density for example $n/n_0 = \Phi(A,B)$ ), this function determines the ratio of similarity, and the quantities A and B are the similar-ity criteria. Shock waves of different GRB (distribution of density, internal energy, etc.) are similar each other if the similarity criteria have the same value. Let us stress, that there are scale-invariant transformations, which leave the similarity criteria unchanged

$$\tilde{E} = k E, \tilde{n}_0 = \lambda n_0, \tilde{r} = (k/\lambda)^{1/3}, \tilde{t} = (k/\lambda)^{1/3} t \qquad (2)$$

Because the observed flux of synchrotron radiation in the optically thin case is the integral of the emissivity over the volume of the shell, «the invariance on the dynamical level leads to scale invariance for the flux within a given spectral regime» (van Eerten & MacFadyen 2012 ). The existence of such symmetry suggests that light curves of similar blast waves can be obtained one from the other by the specified scale transformation. From this it follows that there should be a whole class of objects, whose light curves will have the universal form if time and radiation flux are measured in appropriate units. One can say that a set of such objects constitutes a class of similarity. The universal light curve resting on the self-similar solution bears the imprint of this internal symmetry and therefore can also be symbolically called self-similar.

Let us try to find the above mentioned measure units of time and radiation flux. Suitable choice for time can be found from the following consideration. At the very beginning as the external shock builds up, its bolometric luminosity L rises approximately as $L \sim t^2$ (if a surrounding matter is uniform). At this stage the gamma - factor of the shell remains constant and a radius R is proportional to t. Accordingly, the area of the radiating surface increases as $R^2$. While the mass of the collected gas m is low (m < $E_0 / (\Gamma^2 c^2)$), the luminosity increases. It peaks when R reaches the typical deceleration radius $R_d$, and starts declining rapidly thereafter. The deceleration radius $R_d$ can be easily estimated as follows (Rees & Meszaros 1992): it is the distance where the energy of collected gas becomes comparable to the initial energy of the fireball (the surrounding is supposed to be homogenious):

$$E_0 \approx 4/3 \ \pi \ m_p \ n \ \Gamma^2 \ R_d^3 \qquad (3)$$

Accordingly peak time $T_p$ can be found with the help of the following well known relation connecting R, $\Gamma_0$ and T: $T_p \approx R_d / (4 \Gamma_0^2 c)$. The estimates show also that the typical time scale $T_p$ depends on the same global characteristics that determine self-similar solution as well as radiation properties of the fireball.

As for units of measurement of F, let us remind that the observed flux at a given frequency (i.e. light curve) F_nu can be approximately represented as as a series of power law segments in frequency

$$F_\nu = F_{max} f(\nu,t) \approx Fmax \ (\nu / \nu_p(t))^{-\beta} \qquad (4)$$

Here $\nu_p$ is the peack frequency $\nu_p = min \ ( \nu_m , \nu_c )$ and index β depends on the relation between ν, $\nu_c$ and $\nu_m$ . The critical synchrotron frequency $\nu_m$ is the frequency at which the bulk of the electrons accelerated by the shock radiate, and $\nu_c$ is a cooling frequency - the synchrotron frequency of electrons whose radiative cooling time equals the dynamic timescale (see, for example, Sari et al. 1998). $F_{max}$ and $\nu_p$ change as the power function of time. Hence the flux can be represented as a function of dimensionless variable $t / T_p$ : $F_\nu = F_p \ (t/T_p)^\alpha$ . One can see that a factor $F_p = F_p(T_p)$ is the flux in the maximum of light curve. Thus it is convenient to measure the radiation flux in this units.

From the arguments given above it is clear that with the help of $F_p$ and $T_p$ one can normalize the observed light curves so that curves of different GBR on the graph will be depicted in the same scale. The result of such representation is shown in Fig.2.

The right-hand side of the light curve is naturally described by the standart theory of afterglow (detailed description of the standard model can be found in Piran, 2004) . It is interesting to take a look on the left side of the universal light curve. Here we see a curious coincidence: at small $\tau$, F changes asymptotically as $\tau^2$. The observed universal light curve is changing exactly as bolometric luminosity do it at very early stage of the afterglow. At a later stage according to Gao et al. 2015, the rising slope for $F(\tau)$ ranges between 3 and 1/3 for ISM and wind medium. If one considers the transition between thin and thick shells or a medium index between 0 and 2, the results can be more consistent with the observed data (see Liang et al. 2013 for discussion).

It should be emphasized that Gao et al 2015 did a morphological classification of a wide sample afterglow light curves. It is interesting to note that the set of sources used in this article includes three objects of our sample. All of them are related to one class. The authors also conclude that the optical light curves similar to our universal curve are generated by a single (forward) shock wave.

To summarize, one can say that universal light curve, found in this work, characterizes a whole class of objects for which afterglow is the synchrotron emission of radiatively and adiabatically cooling electrons and continuously injected new electrons in a decelerating forward shock. We should note, that prompt emission can be produced not only from internal shocks, see a detailed discussion of prompt emission mechanism by Zhang 2014.

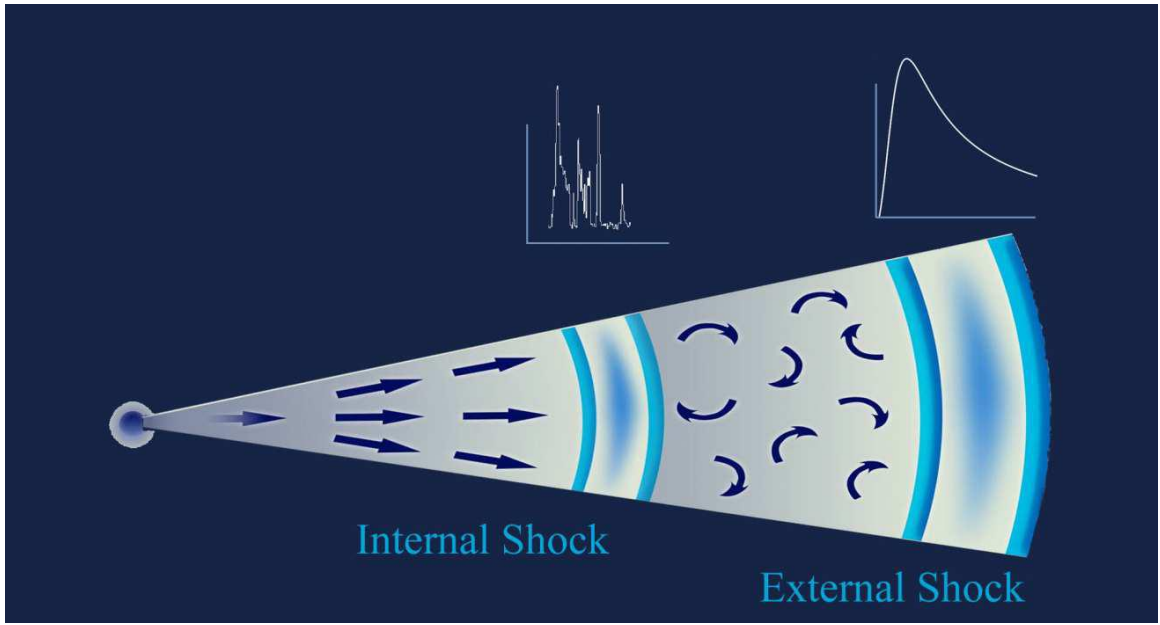

FIGURE 3. Qualitative schematic view of the structure of the relativistic jet produced by the gamma-ray burst. The external shock arises as a result of the impact of the jet on the stellar wind of the progenitor. This is where the final goodbye of the SOS similar emission from the collapsing star forms, which is characterized by a smooth (but non-monotonic) light variation. The internal shock persists as long as the central engine continues operating this is where rapidly varying gamma-, x-ray, and optical radiation forms.

## 6. CONCLUSIONS

Thus the SOS similar emission supports the evidence for a relativistic blast wave plugging into the external medium. The activity in the central region may produce internal shocks that generate gamma-and optical radiation. This radiation is highly variable and shows internal correlations between optical, x-ray, and gamma-ray variations. The external wave (second zone) is actually the stellar wind, which has been strongly compressed by the relativistic shock, and it therefore produces a smoothly and non-monotonically varying optical glow of the self-similar type. This is how SOS similar emission (i.e. the final message of the collapsing matter) forms.

We emphasize that the type of the behavior of the optical emission of gamma-ray bursts discovered in this study does not necessarily mean that it should be exactly reproduced in all cases. Similarly, power-law afterglows show monotonic power-law behavior with a constant slope only as far as average light curves are considered, whereas actual events sometimes may exhibit certain peculiarities like optical flares, etc. SOS emission initial stage may be masked and "broken down" by optical emission correlated with gamma occurs in the internal shock wave. In the case of homogeneities in the density distribution of the stellar wind of progenitors, associated with variations in the physical parameters, the rate of expiration, the magnetic field strength, the exhaust velocity can cause the disappearance of the optical emission. The non-monotonic prompt SOS emission demonstrates the universe self-similar nature of hydrodynamic processes just after its appearance.

## 6. ACKNOWLEDGMENTS

This work was funded in part by the M.V.Lomonosov Moscow State University Development Program, by the Russian Science Foundation agreement 16-12-00085 and by the Russian Foundation of Fundamental Research 15-02-07875 . We thank Dr. Antony Rothman for the fruitfull discussion.